\documentstyle[amssymb,aps,preprint]{revtex}

\begin{document}
\draft
\title{Dynamical content of Chern-Simons Supergravity$\thanks{%
Talk given at the Second Meeting on Trends in Theoretical Physics, Buenos
Aires, November 30-December 4, 1998.}$}
\author{Osvaldo Chand\'{\i }a$^{\natural }$, Ricardo Troncoso and Jorge Zanelli}
\address{Centro de Estudios Cient\'{\i }ficos de Santiago, Casilla 16443, Santiago,
Chile\\
Departamento de F\'{\i }sica, Universidad de Santiago de Chile, Casilla 307,
Santiago 2, Chile\\
$^{\natural }$Instituto de F\'{\i }sica Te\'{o}rica, Rua Pamplona 145, Sao
Paulo, Brazil}
\maketitle

\begin{abstract}
The dynamical content of local $AdS$ supergravity in five dimensions is
discussed. The bosonic sector of the theory contains the vielbein ($e^{a}$),
the spin connection ($\omega ^{ab}$) and internal $SU(N)$ and $U(1)$ gauge
fields. The fermionic fields are complex Dirac spinors ($\psi ^{i}$) in a
vector representation of $SU(N)$. All fields together form a connection
1-form in the superalgebra $SU(2,2|N)$. For $N=4$, the symplectic matrix has
maximal rank in a locally AdS background in which the dynamical degrees of
freedom can be identified. The resulting efective theory have different
numbers of bosonic and fermionic degrees of freedom. 
\end{abstract}

\section{Introduction}

Over twenty years ago, Cremmer, Julia and Scherk presented a beautiful $N=1$
theory of supergravity in 11 dimensions \cite{CJS}, which, apart from the
metric and a gravitino ($\psi $), included a three-form field ($A_{\mu \nu
\lambda }$). This theory is quite unique: a larger $D$ or $N$ would give
rise to inconsistencies upon compactification to 4 dimensions (i.e., fields
with spin greater that 2 or more than one graviton). A ``dual'' alternative
possibility which uses a six form ($A_{\mu _{1}...\mu _{6}}$) instead of the
three form also leads to inconsistencies \cite{NTPvN}. Additionally, it has
been also shown that regardless of the compactification arguments, the
theory cannot accommodate a cosmological constant in eleven dimensions \cite
{BDHS}.

One of the puzzling aspects of the theory is the conjecture contained in the
original paper by Cremmer, Julia and Scherk in the sense that this theory
should be related to another one based on the $osp(32|1)$ algebra, a problem
that they promised to discuss in a forthcoming article that never appeared.
In fact, it can be seen that in a gauge theory for an $osp(32|1)$
connection, the anticommutator of the fermionic generators takes the form%
\cite{AIT} 
\[
\{Q,Q\}\thickapprox P_{a}\Gamma ^{a}+Z_{ab}\Gamma ^{ab}+Z_{abcde}\Gamma
^{abcde}, 
\]
where one could identify the different components of the connection that
accompany the generators in the RHS with the fields in the CJS theory: $%
A_{\mu }^{a}\thicksim e_{\mu }^{a}$, $A_{\mu }^{ab}\thicksim A_{\mu \nu
\lambda }$, $A_{\mu }^{abcde}\thicksim A_{_{\mu _{1}...\mu _{6}}}$. However,
no supergravity theory was known to contain all these fields in a simple and
natural way. It was therefore a surprise for us to find a family of
Lagrangians (one in each odd dimension) which could prove the conjecture of
Cremmer, Julia and Scherk to be true \cite{TrZ}. The key ingredients in this
new family of supergravity theories is their Chern-Simons form and the fact
that the spacetime symmetry is realized in the tangent space and not on the
base manifold.

\section{Supergravity in Odd Dimensions}

One of the unique features of gravity in 2+1 dimensions is that it is a
genuine gauge theory in the sense of having a fiber bundle structure. This
is because the standard Einstein-Hilbert action (both with and without
cosmological constant) is a Chern-Simons ({\bf CS}) form $\int <AdA+\frac{2}{%
3}A^{3}>$. As a consequence, the theory is an integrable system, unlike the
case of four-dimensional gravity \cite{Witten}. Furthermore, the simplest $%
D=3$, $N=1$ supegravity theory \cite{AT} also shares this feature, and as a
bonus, the theory is {\em locally} invariant under the anti-de Sitter group.
It is easy to see that these features can be generalized to all odd
dimensions: provided one has identified the superalgebra that extends AdS in
a given dimension $D$, it is just a matter of constructing the appropriate
Chern-Simons $D$-form. The resulting theory would be invariant by
construction under the right supergroup in which the fields of the theory
transform as different pieces of a connection. Thus, the algebra of the
supersymmetry transformations is guaranteed to close off shell without
requiring auxiliary fields \cite{qmfs6}.

However, both identifying the algebra and the construction of the CS form
involve some subtleties that it is instructive to analyze in detail. The
simplest example that captures all the problems --and which yields a theory
with propagating local degrees of freedom-- occurs in five dimensions.{\bf \ 
}In the following we study the five-dimensional case in detail and indicate
where appropriate how the results generalize to other dimensions.

\section{Higher Dimensional CS Theories}

In 2+1 dimensions, gravity --or any other Chern-Simons theory for that
matter-- has no dynamical degrees of freedom. The field equations are 
\begin{equation}
F=0,  \label{F}
\end{equation}
where $F$ is the (anti-de Sitter or Poincar\'{e}) curvature, and it
therefore means that all on-shell configurations are locally
gauge-equivalent to a flat connection. However, the field equations of a CS
theory in dimensions five and higher --for any gauge group-- take the form

\begin{equation}
<F\mbox{\tiny $\wedge$}F\mbox{\tiny $\wedge$}\cdot \cdot \cdot 
\mbox{\tiny
$\wedge$}F\;G_{M}>=0,  \label{FFF}
\end{equation}
where $G_{M}$ is a gauge generator. This equation in general does not imply
a flat connection, allowing for the existence of propagating degrees of
freedom.

For dimensions $D>3$, the dynamical structure of a CS system becomes highly
nontrivial. The root of the complexity lies in three independent features
inherent of CS theories: {\bf i)} they are gauge systems; {\bf ii)} they
have coordinate invariance built in; and {\bf iii)} they are first order
systems. Although each of these items are neither exclusive of CS systems,
nor particularly difficult to deal with, their conjunction requires special
care. As discussed in \cite{BGH}, there are two problems: diffeomorphism and
gauge invariances are not completely independent, and because of the first
order nature, CS systems have first class constraints inextricably mixed
with second class ones. This makes the Dirac matrix hard to express in a
simple form and almost impossible to invert.

Moreover, for $D>3$ the symplectic form ($\Omega $) that multiplies the
velocities in (\ref{FFF}) is a function of the gauge field $\Omega =\Omega (%
{\bf A})$, and it can degenerate for certain field configurations. In
particular, for $D=5$, 
\begin{equation}
\Omega _{MN}^{ij}({\bf A})=\Delta _{MNP}\epsilon ^{ijkl}{\bf F}_{kl}^{P},
\end{equation}
where $\Delta _{MNP}\equiv <{\bf G}_{M}{\bf G}_{N}{\bf G}_{P}>$ is an
invariant tensor of the gauge group (see Appendix B). At the singular
configurations (as ${\bf F}_{kl}^{P}=0$), the rank of the Dirac matrix is
reduced and becomes noninvertible, so that some (or all) second class
constraints could actually be viewed as first class, and this further
complicates the identification of the propagating degrees of freedom.
Outside those regions, however, other dynamical structure is well behaved
and the symplectic form has maximal rank.

Here we will not discuss the problems arising from the presence of
degenerate surfaces. This seems to be a reasonable point of view as these
singularities constitute sets of measure zero in the configuration space of
the theory.

The dynamical analysis of a higher-dimensional bosonic CS theory is
discussed in \cite{BGH} and we summarize it here: If the gauge algebra takes
the form $G\times U(1)$, where $G$ is a semisimple algebra, the symplectic
form can be computed in a background where $\Omega $ has maximal rank. The
configurations that satisfy this requirement are called generic in the sense
that under small deformations the rank of $\Omega $ remains maximal. Also,
in these configurations, the first and second class constraints can be
separated and the degrees of freedom computed. If the algebra has dimension $%
f>1$ and the spacetime has dimension $D=2n+1>3$, it is shown that the number
of propagating local degrees of freedom of the theory is 
\[
g=nf-f-n 
\]
and, in five dimensions, the symplectic matrix has the form 
\begin{equation}
\Omega _{MN}^{ij}({\bf \bar{A}})=g_{MN}\epsilon ^{ijkl}{\bf \bar{f}}_{kl},
\end{equation}
where the bar stands for background fields, and ${\bf \bar{f}}_{kl}=\partial
_{k}\bar{b}_{l}-$ $\partial _{l}\bar{b}_{k}$ is the curvature of the $U(1)$
field.

\section{Local AdS$_{5}$ Supergravity}

The supersymmetric extension of the anti-de Sitter algebra in five
dimensions is $su(2,2|N)$ \cite{Nahm}, whose associated connection can be
written as, 
\[
{\bf A=}e^{a}{\bf J}_{a}+\frac{1}{2}\omega ^{ab}{\bf J}_{ab}+a^{\Lambda }%
{\bf T}_{\Lambda }+(\bar{\psi}^{r}{\bf Q}_{r}-{\bf \bar{Q}}^{r}\psi _{r})+b%
{\bf Z}, 
\]
where the generators ${\bf J}_{a}$, ${\bf J}_{ab}$, form an AdS algebra ($%
so(4,2)$), ${\bf T}_{\Lambda }$ ($\Lambda =1,\cdot \cdot \cdot N^{2}-1$) are
the generators of $su(N)$, ${\bf Z}$ generates a $U(1)$ subgroup and ${\bf Q}%
,{\bf \bar{Q}}$ are the supersymmetry generators, which transform in a
vector representation of $SU(N)$. The Chern-Simons Lagrangian for this gauge
algebra is defined by the relation $dL=i<{\bf F}^{3}>$, where ${\bf F=dA+}$ $%
{\bf A}^{2}${\bf \ }is the (antihermitean) curvature, and $<\cdot \cdot
\cdot >$ stands for the supertrace in the representation described in the
Appendix A. Using this definition, one obtains the Lagrangian originally
discussed by Chamseddine in \cite{Chams},

\begin{equation}
L=L_{G}(\omega ^{ab},e^{a})+L_{su(N)}(a_{s}^{r})+L_{u(1)}(\omega
^{ab},e^{a},b)+L_{F}(\omega ^{ab},e^{a},a_{s}^{r},b,\psi _{r}),  \label{L}
\end{equation}
with

\begin{equation}
\begin{array}{lll}
L_{G} & = & \frac{1}{8}\epsilon _{abcde}\left[ R^{ab}R^{cd}e^{e}/l+\frac{2}{3%
}R^{ab}e^{c}e^{d}e^{e}/l^{3}+\frac{1}{5}e^{a}e^{b}e^{c}e^{d}e^{e}/l^{5}%
\right] \\ 
L_{su(N)} & = & -Tr\left[ a(da)^{2}+\frac{3}{2}a^{3}da+\frac{3}{5}%
a^{5}\right] \\ 
L_{u(1)} & = & \left( \frac{1}{4}-\frac{1}{N}\right) b(db)^{2}+\frac{3}{%
4l^{2}}\left[ T^{a}T_{a}-R^{ab}e_{a}e_{b}-l^{2}R^{ab}R_{ab}/2\right] b \\ 
&  & +\frac{3}{N}f_{s}^{r}f_{r}^{s}b \\ 
L_{F} & = & 
\begin{array}{l}
\frac{3}{2i}\left[ \bar{\psi}^{r}{\cal R}\nabla \psi _{r}+\bar{\psi}^{s}%
{\cal F}_{s}^{r}\nabla \psi _{r}\right] +c.c.
\end{array}
\end{array}
,  \label{Li}
\end{equation}
where $a_{s}^{r}\equiv a^{\Lambda }(\tau _{\Lambda })_{s}^{r}$ is the $%
su(2,2)$ connection, $f_{s}^{r}$ is its curvature, and the bosonic blocks of
the supercurvature: ${\cal R}=\frac{1}{2}T^{a}\Gamma _{a}+\frac{1}{4}%
(R^{ab}+e^{a}e^{b})\Gamma _{ab}+\frac{i}{4}dbI-\frac{1}{2}\psi _{s}\bar{\psi}%
^{s}$, ${\cal F}_{s}^{r}=f_{s}^{r}+\frac{i}{N}db\delta _{s}^{r}-\frac{1}{2}%
\bar{\psi}^{r}\psi _{s}$. The cosmological constant is $-l^{-2},$ and the
AdS covariant derivative $\nabla $ acting on $\psi _{r}$ is

\begin{equation}
\nabla \psi _{r}=D\psi _{r}+\frac{1}{2l}e^{a}\Gamma _{a}\psi
_{r}-a_{\,r}^{s}\psi _{s}+i\left( \frac{1}{4}-\frac{1}{N}\right) b\psi _{r}.
\label{delta}
\end{equation}
where $D$ is the Lorentz covariant derivative.

The above relation implies that the fermions carry a $u(1)$ ``electric''
charge given by $e=\left( \frac{1}{4}-\frac{1}{N}\right) $. The purely
gravitational part, $L_{G}$ is equal to the standard Einstein-Hilbert action
with cosmological constant, plus the dimensionally continued Euler density%
\footnote{%
The first term in $L_{G}$ is the dimensional continuation of the Euler (or
Gauss-Bonnet) density from two and four dimensions, exactly as the
three-dimensional Einstein-Hilbert Lagrangian is the continuation of the the
two dimensional Euler density. This is the leading term in the limit of
vanishing cosmological constant ($l\rightarrow \infty )$, whose local
supersymmetric extension yields a nontrivial extension of the Poincar\'{e}
group \cite{btrz}.}.

The action is by construction invariant --up to a surface term-- under the
local (gauge generated) supersymmetry transformations $\delta _{\lambda
}A=-(d\lambda +[A,\lambda ])$ with $\lambda =\bar{\epsilon}^{r}{\bf Q}_{r}-%
{\bf \bar{Q}}^{r}\epsilon _{r}$, or

\[
\begin{array}{lll}
\delta e^{a} & = & \frac{1}{2}\left( \overline{\epsilon }^{r}\Gamma ^{a}\psi
_{r}-\bar{\psi}^{r}\Gamma ^{a}\epsilon _{r}\right) \\ 
\delta \omega ^{ab} & = & -\frac{1}{4}\left( \bar{\epsilon}^{r}\Gamma
^{ab}\psi _{r}-\bar{\psi}^{r}\Gamma ^{ab}\epsilon _{r}\right) \\ 
\delta a_{\,s}^{r} & = & -i\left( \bar{\epsilon}^{r}\psi _{s}-\bar{\psi}%
^{r}\epsilon _{s}\right) \\ 
\delta \psi _{r} & = & -\nabla \epsilon _{r} \\ 
\delta \bar{\psi}^{r} & = & -\nabla \bar{\epsilon}^{r} \\ 
\delta b & = & -i\left( \bar{\epsilon}^{r}\psi _{r}-\bar{\psi}^{r}\epsilon
_{r}\right) .
\end{array}
\]

As can be seen from (\ref{Li}) and (\ref{delta}), for $N=4$ the $b$ field
looses its kinetic term and decouples from the fermions (the gravitino
becomes uncharged with respect to the $U(1)$ field). The only remnant of the
interaction with the $b$ field is a dilaton-like coupling with the
Pontryagin four forms for the AdS and $SU(N)$ groups (in the bosonic
sector). As it is also shown in the Appendix A, the case $N=4$ is also
special at the level of the algebra, which becomes a superalgebra with a $%
u(1)$ central extension.

In the bosonic sector, for $N=4$, the field equation obtained from the
variation with respect to $b$ states that the Pontryagin four form of AdS
and $SU(N)$ groups are proportional . Consequently, if the curvatures
approach zero sufficiently fast at spatial infinity, there is a conserved
topological current which states that, for the spatial section, the second
Chern characters of AdS\ and $SU(4)$ are proportional. Consequently, if the
spatial section has no boundary, the corresponding Chern numbers are
related. Using the fact that $\Pi _{4}(SU(4))=0$, the above implies that the
Hirzebruch signature plus the Nieh-Yan number of the spatial section cannot
change in time.

\section{Symplectic form}

We will consider a background that is a solution for the field equations,
for which $\Omega $ has maximum rank. Realizing this is in general a
difficult task. However, an amazing simplification happen when $N=4$, which
has its root in the form of the invariant tensor $\Delta _{MNP}$. As stated
in the Appendix, considering the spliting: $M=\left\{ M^{\prime },Z\right\} $%
, when $N=4$ we have ``the accident'': $\Delta _{ZZZ}=0$, and $\Delta
_{ZM^{\prime }N^{\prime }}=-\frac{i}{4}g_{M^{\prime }N^{\prime }}$ , where $%
g_{M^{\prime }N^{\prime }}$ is the Killing metric of $PSU(2,2|4)$.

This fact will help us find an adequate background. Consider any locally AdS
spacetime with pure gauge matter fields (Bosons and fermions), with the
exception of the $b$ field. That is, a background ${\bf \bar{A}}$ such that $%
{\bf F}^{AB}={\bf F}_{\Lambda }={\bf \psi }_{s}{\bf =\bar{\psi}}^{r}=0\neq 
{\bf F}^{Z}$.

It is easy to see also that for $N=4$ the conditions that make the
separation between first and second class constraints possible in the
construction of \cite{BGH} can also be applied, even if in this case the
algebra is bigraded and {\em not} a direct sum of a semisimple one and $u(1)$%
, but a central extension.

Indeed, it can be directly checked that the symplectic form takes the form

\begin{equation}
\Omega _{MN}^{ij}\left[ {\bf \bar{A}}\right] =\left[ 
\begin{array}{cc}
0_{4\times 4} & 0 \\ 
0 & \widehat{\Omega }_{M^{\prime }N^{\prime }}^{ij}
\end{array}
\right]  \label{Omega}
\end{equation}
where $\widehat{\Omega }_{M^{\prime }N^{\prime }}^{ij}$ is generically an
invertible matrix. In fact, for a flat AdS curvature (e.g., a spacetime of
constant negative curvature and vanishing torsion) the symplectic matrix
takes the form

\begin{equation}
\Omega _{MN}^{ij}\left[ {\bf \bar{A}}\right] =\left[ 
\begin{array}{ccccc}
0 & 0 & 0 & 0 & 0 \\ 
0 & \eta _{_{[AB][CD]}} & 0 & 0 & 0 \\ 
0 & 0 & g_{_{\Lambda _{1}\Lambda _{2}}} & 0 & 0 \\ 
0 & 0 & 0 & 0 & 2\delta _{n}^{r}\delta _{\alpha }^{\beta } \\ 
0 & 0 & 0 & -2\delta _{s}^{m}\delta _{\beta }^{\alpha } & 0
\end{array}
\right] \otimes -\frac{i}{4}\epsilon ^{ijkl}(\partial _{k}\bar{b}%
_{l}-\partial _{l}\bar{b}_{k}).  \label{Omega'}
\end{equation}

The nonvanishing block in the RHS can be recognized as the Killing metric
for $PSU(2,2|4)$, while the factor on the right is the space-dual of the $b$
field, $*db$.

This shows that the requirement for the algebra to be of the form $G\times
U(1)$ in order to decouple first and second class constraints is sufficient
but not necessary. In general, any semisimple (super) group with a $U(1)$
central extension [as in case of $N=4$ super AdS$_{5}$ discussed above] will
be sufficient too.

One can now count the degrees of freedom for the effective theory in this
background. There are $15$ generators of AdS$_{5}$, $15$ for $SU(2,2)$, and
1 for $U(1)$, plus $4\times 4$ $Q_{\alpha }^{i}$'s and an equal number of $%
\bar{Q}_{i}^{\alpha }$ 's. According to the argument outlined in \cite{BGH},
this makes $f=63$, $n=2$, and a total of 61 independently propagating
degrees of freedom.

The previous result is puzzling. There can be no matching of fermionic and
bosonic degrees of freedom in this case. In fact, there seems to be a hidden
subtlety in this counting because, at least in the perturbative sense, the
number of degrees of freedom around this background is different. Consider a
fluctuation in the connection around a fixed background ${\bf \bar{A}}$ , 
\begin{equation}
{\bf A}^{M}={\bf \bar{A}}^{M}+{\bf u}^{M},
\end{equation}
where the dynamical fields will be the spatial components of ${\bf A=A}^{M}%
{\bf G}_{M}$, with $M$ ranging over the whole supergroup indices. Then, for
small ${\bf u}$, the effective action in the linearized approximation, is
given by

\begin{eqnarray}
{\bf I}_{Eff}({\bf u}_{i}^{M}) &\sim &\int <{\bf u\bar{F}}\bar{\nabla}{\bf u}%
>  \nonumber \\
&=&\int d^{5}x\;[{\bf u}_{i}^{M}\Omega _{MN}^{ij}({\bf \bar{A}})\bar{\nabla}%
_{0}{\bf u}_{j}^{N}+2{\bf u}_{0}^{M}\Omega _{MN}^{ij}({\bf \bar{A}})\bar{%
\nabla}_{i}{\bf u}_{j}^{N}-2\varepsilon ^{ijkl}{\bf u}_{i}^{A}\Delta _{ABC}%
{\bf \bar{F}}_{oj}^{B}\bar{\nabla}_{k}{\bf u}_{l}^{C}].  \label{Lpert}
\end{eqnarray}
If we consider the AdS background where $\Omega _{MN}^{ij}({\bf \bar{A}})$
takes the form (\ref{Omega}), then the counting comes to 58.

\section{Discussion and Outlook}

{\bf 1.} From the previous discussion it is clear that for $N=4$ the theory
is extremely simple and the $b$ field almost completely decouples from the
rest of the fields. In fact, the $b$ field is analogous to a Lagrange
multiplier, and this analogy is completely accurate in the effective action
for the perturbations, where the perturbation associated with this field
doesn't appear at all in the effective action. Actually, around the same AdS
background one cannot distinguish between the effective linearized theory
described above from that containing only the K\"{a}hler-CS like action \cite
{Nair} described by the second term of $L_{u(1)\text{ }}$.{\bf \ }We will
discuss this issue in detail elsewhere.

{\bf 2.} Another important problem is to show that the algebra generated by
the conserved charges reproduces an algebra which {\em is not} isomorphic to
the gauge algebra $su(2,2|4)$, but is a non-trivial central extension of $%
psu(2,2|4)$. Here the result is that this is indeed the case and that the
algebra sets bounds to the values of the charges and conditions for the
existence of Killing spinors that ensure that the background saturates the
Bogomolny'i bound.{\bf \ }This issue, in turn, raises the question of
identifying the nontrivial BPS states. These states must keep a fraction of
supersymmetry; in fact a solution of the field equations with these features
is some sort of ``topological black hole'' \cite{BGM} and it is reassuring
to verify that the extreme case saturates the bound. These problems are
going to be discussed in a forthcoming article.

{\bf 3}. The five dimensional theory generates a new four-dimensional
superconformal theory at the boundary. This theory is constructed on the
generalization of the Ka\v {c}-Moody extension of the superconformal algebra
without central charge, that is, the WZW$_{4}$ algebra for $PSU(2,2|1)$.
This theory could be a rich test ground in the context of AdS/CFT duality
conjecture \cite{Maldacena}.

{\bf 4.} Another interesting problem is try to generalize what is known
about $D=5$ to higher dimensions. We have shown \cite{TrZ} that the $D=11$, $%
N=32$ theory admits a nontrivial extension of the AdS superalgebra with one
abelian generator for which anti-de Sitter space without matter fields is a
background of maximal rank, and the gauge superalgebra is realized in the
Dirac brackets. On a background like the used in the five dimensional
theory, the $D=11$ theory has $2^{12}$ fermionic and $2^{12}-1$ bosonic
degrees of freedom, and the (super) charges obey a non-trivial central
extension of the $OSP(32|32)$ algebra.

\noindent {\large {\bf Acknowledgments}}\newline

The authors are grateful to R. Aros, M. Contreras, J. Gamboa, J. M. F.
Labastida, J. Maldacena, C. Mart\'{\i }nez, F. M\'{e}ndez, P. van
Nieuwenhuizen, R. Olea, C. Teitelboim for many enlightening discussions and
helpful comments. This work was supported in part through grants 1990189,
1970151, 1980788, and 3960009 from FONDECYT, and grant 27-953/ZI-DICYT
(USACH). The institutional support of FUERZA AEREA DE CHILE, I.\
Municipalidad de Las Condes, and a group of Chilean companies (AFP
Protecci\'{o}n, Business Design Associates, CGE, CODELCO, COPEC, Empresas
CMPC, GENER\ S.A., Minera Collahuasi, Minera Escondida, NOVAGAS and
XEROX-Chile) is also recognized. R.T and J.Z. wish to express their
gratitude to Marc Henneaux for his kind hospitality in Brussels and for many
fruitful discussions and key comments. Last but not least, we wish to thank
the organizers for the warm atmosphere at the meeting in Buenos Aires.

\section{Appendices}

\appendix

\section{Supersymmetric extension of AdS$_{5}\;$algebra}

As discussed in \cite{Nahm}, \cite{qmfs6}, the supersymmetric extension of
the anti-de Sitter algebra in five dimensions is $su(2,2|N)$. This is the
Lie algebra associated with the invariance group of the quadratic form $%
q=\theta ^{*\alpha }g_{\alpha \beta }\theta ^{\beta }+z^{*r}u_{rs}z^{s}$,
with $\alpha ,\beta =1,...,4$ and $r,s=1,...,N$. Here $\theta ^{\alpha }$
are complex Grassmann numbers and $g_{\alpha \beta }$ and $u_{pq}$ are
sesquilinear metrics, which will be chosen as $g_{\alpha \beta }=i(\gamma
_{0})_{\alpha \beta }$ and $u_{rs}=\delta _{rs}$. The supersymmetric algebra
contains $su(2,2)\oplus su(N)\oplus u(1)$ as the bosonic subalgebra. Using
the isomorphism: $su(2,2)\simeq so(4,2)$, the generators are composed by the
AdS generators (${\bf J}_{AB}$), with $A,B=0,...5$, the $2\times 4N$
(complex) supersymmetry generators (${\bf Q}_{r}^{\alpha }$, ${\bf \bar{Q}}%
_{\alpha }^{r}$), and the rest of the algebra is composed by the generators
of internal (Lorentz scalar) $su(N)\otimes u(1)$ (${\bf T}_{\Lambda }$, $%
{\bf Z}$), with $\Lambda =1,...,N^{2}-1$.

A natural representation acting in the vector superspace $(\theta ^{\beta }$%
; $z^{q})$ is given by $\!(4+N)\times (4+N)$ matrices as follows. Defining 
{\bf I}$_{4\times 4}$=$\left[ 
\begin{array}{ll}
\delta _{\beta }^{\alpha } & 0 \\ 
0 & 0
\end{array}
\right] $, {\bf I}$_{N\times N}$ = $\left[ 
\begin{array}{ll}
0 & 0 \\ 
0 & \delta _{s}^{r}
\end{array}
\right] $, the generators are:

$\bullet ${\bf AdS generators}

\begin{eqnarray}
{\bf J}_{AB} &=&\frac{1}{2}\Gamma _{AB}\otimes {\bf I}_{4\times 4}, \\
&=&\left[ 
\begin{array}{cc}
\frac{1}{2}(\Gamma _{AB})_{\beta }^{\alpha } & 0 \\ 
0 & 0
\end{array}
\right] ,  \label{ads}
\end{eqnarray}

$\bullet $ $su(N)${\bf \ generators}

\begin{eqnarray}
{\bf T}_{\Lambda } &=&\tau _{\Lambda }\otimes {\bf I}_{N\times N}  \label{T}
\\
&=&\left[ 
\begin{array}{cc}
0 & 0 \\ 
0 & (\tau _{\Lambda })_{s}^{r}
\end{array}
\right]  \label{sun}
\end{eqnarray}
where $(\tau _{\Lambda })_{s}^{r}$ are the antihermitean generators of $%
su(n) $.

$\bullet ${\bf Supersymmetry generators} 
\begin{eqnarray*}
{\bf Q}_{s}^{\alpha } &=&\left[ 
\begin{array}{cc}
0 & 0 \\ 
-\delta _{\beta }^{\alpha }\delta _{s}^{r} & 0
\end{array}
\right] , \\
{\bf \bar{Q}}_{\alpha }^{r} &=&\left[ 
\begin{array}{cc}
0 & \,\;\delta _{q}^{r}\delta _{\alpha }^{\beta } \\ 
0 & 0
\end{array}
\right] .
\end{eqnarray*}

$\bullet $ $u(1)${\bf \ charge}

\begin{eqnarray}
{\bf Z} &=&\frac{i}{4}{\bf I}_{B}+\frac{i}{N}{\bf I}_{F}  \label{zeta} \\
&=&\left[ 
\begin{array}{ll}
\frac{i}{4}\delta _{\beta }^{\alpha } & 0 \\ 
0 & \frac{i}{N}\delta _{s}^{r}
\end{array}
\right] .
\end{eqnarray}
The commutators of the bosonic generators are those for the algebra $%
so(4,2)\oplus su(N)\oplus u(1)$: $
\begin{array}{lll}
\left[ {\bf J,J}\right] & \sim & {\bf J}
\end{array}
$, $
\begin{array}{lll}
\left[ {\bf T,T}\right] & \sim & {\bf T}
\end{array}
$, $
\begin{array}{lll}
\left[ {\bf Z,Z}\right] & = & 0
\end{array}
$, $
\begin{array}{lll}
\left[ {\bf J,T}\right] & = & 0
\end{array}
$, $
\begin{array}{lll}
\left[ {\bf Z,J}\right] & = & 0
\end{array}
$, $
\begin{array}{lll}
\left[ {\bf Z,T}\right] & = & 0
\end{array}
$. The supersymmetry generators transforms as spinors under AdS (then also
under Lorentz), as ``vectors'' under $su(N)$, and carry $u(1)$ charge, 
\begin{equation}
\begin{array}{lll}
\left[ {\bf J}_{AB}{\bf ,Q}_{s}^{\alpha }\right] & = & -\frac{1}{2}({\bf %
\Gamma }_{AB})_{\beta }^{\alpha }{\bf Q}_{s}^{\beta } \\ 
\left[ {\bf J}_{AB}{\bf ,\bar{Q}}_{\beta }^{r}\right] & = & \;\,\,\frac{1}{2}%
({\bf \Gamma }_{AB})_{\beta }^{\alpha }{\bf \bar{Q}}_{\alpha } \\ 
\left[ {\bf T}_{\Lambda }{\bf ,Q}_{s}^{\alpha }\right] & = & \;\,\,(\tau
_{\Lambda })_{s}^{r}{\bf Q}_{r}^{\alpha } \\ 
\left[ {\bf T}_{\Lambda }{\bf ,\bar{Q}}_{\beta }^{r}\right] & = & -(\tau
_{\Lambda })_{s}^{r}{\bf \bar{Q}}_{\beta }^{s} \\ 
\left[ {\bf Z,Q}_{s}^{\alpha }\right] & = & -i(\frac{1}{4}-\frac{1}{N}){\bf Q%
}_{s}^{\alpha } \\ 
\left[ {\bf Z,\bar{Q}}_{\beta }^{r}\right] & = & \,\;i(\frac{1}{4}-\frac{1}{N%
}){\bf \bar{Q}}_{\beta }^{r}.
\end{array}
\label{BF}
\end{equation}

Finally, the anticommutator reads

\begin{equation}
\left\{ {\bf Q}_{s}^{\alpha },\overline{{\bf Q}}_{\beta }^{r}\right\} =-%
\frac{1}{2}\delta _{s}^{r}(\Gamma ^{a})_{\beta }^{\alpha }{\bf J}_{a}+\frac{1%
}{4}\delta _{s}^{r}(\Gamma ^{ab})_{\beta }^{\alpha }{\bf J}_{ab}-\delta
_{\beta }^{\alpha }({\bf \tau }^{\Lambda })_{\,s}^{r}T_{\Lambda }+i\delta
_{\beta }^{\alpha }\delta _{s}^{r}{\bf Z}.  \label{QQ}
\end{equation}
where ${\bf J}_{a}:={\bf J}_{a5}$.

It is clear from the algebra that the case $N=4$ is a special one: the
generator ${\bf Z}$ commutes with the rest of the algebra and it is just a
central extension, as can be read from the right hand of (\ref{QQ}).

It is important to point out that, if ${\bf Z}$ were omitted, the new
algebra, $psu(2,2|4)$, does not admit the representation described above,
but still satisfies the Jacobi identity. Because of this last feature, the
full resulting algebra for $N=4$, {\bf is not }$psu(2,2|4)\oplus u(1)$.

\section{Third rank invariant tensor for N extended super AdS$_{5}$}

From the matrix representation described above, it is straighforward to
obtain a third rank invariant tensor for the AdS$_{5}$ supergroup, which is
needed to analyze the dynamics of local AdS$_{5}$ supergravity.

Let ${\bf G}_{M}$ the generators of $su(2,2|N)$, where the index $M$ ranks
from the whole superalgebra: $M=\left\{ [AB],\Lambda ,Z,\left( _{\alpha
}^{r}\right) ,\left( _{s}^{\beta }\right) \right\} $.

The required tensor is defined through $\Delta _{MNP}$ = $<{\bf G}_{M},{\bf G%
}_{N},{\bf G}_{P}>$, where $<...>$ stands for the supertrace, which ensure
the invariance of the tensor under the action of the group. Because the
supertrace is the difference between the trace of the upper and lower
diagonal bosonic blocks, the invariant tensor has the form:

\[
\begin{array}{lll}
\Delta _{ZZZ} & = & -i\left( \frac{1}{4^{2}}-\frac{1}{N^{2}}\right) \\ 
\Delta _{[AB][CD][EF]} & = & \frac{i}{2}{\large \epsilon }_{ABCDEF} \\ 
\Delta _{\left( \Lambda _{1}\right) \left( \Lambda _{2}\right) \left(
\Lambda _{3}\right) } & = & -Tr\left[ (\tau _{\Lambda _{1}})(\tau _{\Lambda
_{2}})(\tau _{\Lambda _{3}})\right] \\ 
\Delta _{Z[AB][CD]} & = & -\frac{i}{4}{\Large \eta }_{_{[AB][CD]}} \\ 
\Delta _{Z\left( \Lambda _{1}\right) \left( \Lambda _{2}\right) } & = & -%
\frac{i}{N}g_{_{\Lambda _{1}\Lambda _{2}}} \\ 
\Delta _{Z\left( _{s}^{\beta }\right) \left( _{\alpha }^{r}\right) } & = & 
i\left( \frac{1}{4}+\frac{1}{N}\right) \delta _{\alpha }^{\beta }\delta
_{s}^{r} \\ 
\Delta _{[AB]\left( _{s}^{\beta }\right) \left( _{\alpha }^{r}\right) } & =
& \frac{1}{2}(\Gamma _{AB})_{\alpha }^{\beta }\delta _{s}^{r} \\ 
\Delta _{\left( \Lambda \right) \left( _{s}^{\beta }\right) \left( _{\alpha
}^{r}\right) } & = & -\delta _{\alpha }^{\beta }(\tau _{\Lambda })_{s}^{r},
\end{array}
\]
where $\eta _{_{[AB][CD]}}:=\eta _{_{AC}}\eta _{_{BD}}-\eta _{_{AD}}\eta
_{_{BC}})$, and $g_{_{\Lambda _{1}\Lambda _{2}}}=Tr\left[ (\tau _{_{\Lambda
_{1}}})(\tau _{_{\Lambda _{2}}})\right] $ is the Killing metric of $su(N)$.

Note that, again in the special case $N=4$, and considering the splitting $%
M=\left\{ M^{\prime },Z\right\} $, then $\Delta _{ZZZ}=0$, and $\Delta
_{ZM^{\prime }N^{\prime }}=-\frac{i}{4}g_{M^{\prime }N^{\prime }}$ ,where $%
g_{M^{\prime }N^{\prime }}$ is the (invertible) Killing metric of $%
PSU(2,2|4) $.


\end{document}